\documentclass[journal=ancac3,manuscript=article]{achemso}
\setkeys{acs}{articletitle=true,etalmode=truncate,maxauthors=20}

\usepackage[version=3]{mhchem} % Formula subscripts using \ce{}
\usepackage[T1]{fontenc}       % Use modern font encodings
\usepackage{graphicx}
\usepackage{color}
\usepackage{float}
\usepackage[utf8x]{inputenc}

\usepackage{pdfpages}

\author{G. Kremer}
\affiliation[A]{D{\'e}partement de Physique and Fribourg Center for Nanomaterials, Universit{\'e} de Fribourg, CH-1700 Fribourg, Switzerland}
\alsoaffiliation[B]{Universit{\'e} Paris-Saclay, CNRS, Centre de Nanosciences et de Nanotechnologies, 91120, Palaiseau, France}

\author{J. Maklar}
\affiliation{Fritz Haber Institute of the Max Planck Society, Faradayweg 4-6, 14195 Berlin, Germany}

\author{L. Nicolaï}
\affiliation{New Technologies-Research Center University of West Bohemia, Plzen, Czech Republic}

\author{C. W. Nicholson}
\affiliation[A]{D{\'e}partement de Physique and Fribourg Center for Nanomaterials, Universit{\'e} de Fribourg, CH-1700 Fribourg, Switzerland}
\alsoaffiliation{Fritz Haber Institute of the Max Planck Society, Faradayweg 4-6, 14195 Berlin, Germany}

\author{C. Yue}
\affiliation[A]{D{\'e}partement de Physique and Fribourg Center for Nanomaterials, Universit{\'e} de Fribourg, CH-1700 Fribourg, Switzerland}

\author{C. Silva}
\affiliation{Fritz Haber Institute of the Max Planck Society, Faradayweg 4-6, 14195 Berlin, Germany}

\author{P. Werner}
\affiliation[A]{D{\'e}partement de Physique and Fribourg Center for Nanomaterials, Universit{\'e} de Fribourg, CH-1700 Fribourg, Switzerland}

\author{J. H. Dil}
\affiliation{Photon Science Division, Paul Scherrer Institut, CH-5232 Villigen, Switzerland}
\alsoaffiliation{Institute of physics, Ecole Polytechnique F{\'e}d{\'e}rale de Lausanne, CH-1015 Lausanne, Switzerland}

\author{J. Krempask{\'y}}
\affiliation{Photon Science Division, Paul Scherrer Institut, CH-5232 Villigen, Switzerland}

\author{G. Springholz}
\affiliation{Institut f{\"u}r Halbleiter-und Festk{\"o}rperphysik, Johannes Kepler Universit{\"a}t, A-4040 Linz, Austria}

\author{R. Ernstorfer}
\affiliation{Fritz Haber Institute of the Max Planck Society, Faradayweg 4-6, 14195 Berlin, Germany}
\alsoaffiliation{Institut für Optik und Atomare Physik, Technische Universität Berlin, Stra{\ss}e des 17, Juni 135, 10632 Berlin, Germany}

\author{J. Min{\'ar}}
\affiliation{New Technologies-Research Center University of West Bohemia, Plzen, Czech Republic}

\author{L. Rettig}
\affiliation{Fritz Haber Institute of the Max Planck Society, Faradayweg 4-6, 14195 Berlin, Germany}

\author{C. Monney}
\affiliation[A]{D{\'e}partement de Physique and Fribourg Center for Nanomaterials, Universit{\'e} de Fribourg, CH-1700 Fribourg, Switzerland}

\title[]
  {Field-induced ultrafast modulation of Rashba coupling at room temperature in ferroelectric $\alpha$-GeTe(111)}

\begin{document}

%\linenumbers

\newpage

\textbf{Rashba materials have appeared as an ideal playground for spin-to-charge conversion in prototype spintronics devices\cite{sanchez2013spin,lesne2016highly}. Among them, $\alpha$-GeTe(111) is a non-centrosymmetric ferroelectric (FE) semiconductor for which a strong spin-orbit interaction gives rise to giant Rashba coupling\cite{di2013electric,liebmann2016giant,krempasky2016disentangling,rinaldi2018ferroelectric,PhysResearchKrempasky2020,kremer2020}. 
Its room temperature ferroelectricity was recently demonstrated as a route towards a new type of highly energy-efficient non-volatile memory device based on switchable polarization \cite{varotto2021room}. Currently based on the application of an electric field \cite{noel2020non,varotto2021room}, the writing and reading processes could be outperformed by the use of femtosecond (fs) light pulses requiring exploration of the possible control of ferroelectricity on this timescale. Here, we probe the room temperature transient dynamics of the electronic band structure of $\alpha$-GeTe(111) using time and angle-resolved photoemission spectroscopy \hbox{(tr-ARPES).} Our experiments reveal an ultrafast modulation of the Rashba coupling mediated on the fs timescale by a surface photovoltage (SPV), namely an increase corresponding to a \hbox{13 $\%$} enhancement of the lattice distortion. This opens the route for the control of the FE polarization in $\alpha$-GeTe(111) and FE semiconducting materials in quantum heterostructures.}

\bigskip

\begin{figure}[H]
\begin{center}
\includegraphics[width=100mm]{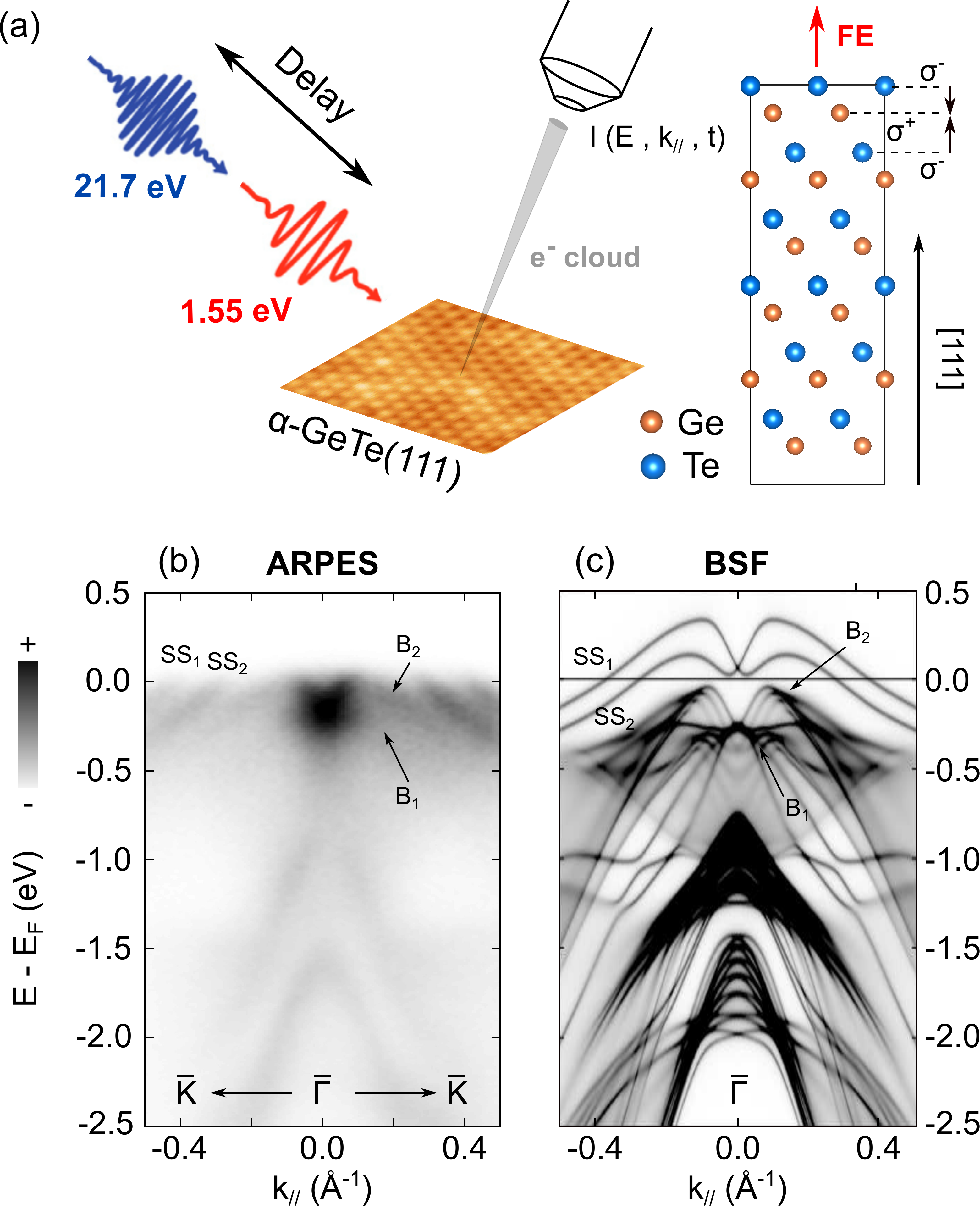}
\end{center}
\caption{\label{fig1} \textbf{tr-ARPES configuration and static ARPES characterization.} (a) tr-ARPES experiment using $21.7$ eV probe and $1.55$ eV pump pulses (left), and surface structure of $\alpha$-GeTe along the [111] crystallographic direction with Te termination and short surface bonds (right). Interatomic dipoles (black arrows) due to negative ($\sigma^{-}$) and positive ($\sigma^{+}$) net charges give rise to a net outward FE polarization (red arrow). (b) Static ARPES spectrum along the $\bar{K}-\bar{\Gamma}-\bar{K}$ high-symmetry direction (see second derivative in Fig. S1) and (c) corresponding BSF calculations for a Te-terminated surface of  $\alpha$-GeTe(111) with short surface bonds.}
\end{figure}

Intense fs light pulses present a powerful tool to control the physical properties of solids, promoting new emergent phenomena inaccessible in the ground state, \cite{stojchevska2014ultrafast,kogar2020light,maklar2021nonequilibrium,beaulieu2021ultrafast} as evidenced in FE materials such as perovskites \cite{nova2019metastable}. In the familly of ferroelectrics, \hbox{$\alpha$-GeTe(111)} is a fascinating material because it  exhibits a giant FE distortion below $T_{C} \approx 700 \thinspace K$, leading to spin-polarized bulk and surface-split electronic states with the largest Rashba parameter so far reported \cite{di2013electric,liebmann2016giant,krempasky2016disentangling}. Based on the inverse spin Hall and inverse Rashba-Edelstein effects, Rashba systems can be efficiently used for spin-to-charge conversion in spintronics devices \cite{sanchez2013spin,lesne2016highly,varotto2021room}. \hbox{$\alpha$-GeTe(111)} has been identified as a promising candidate in that direction but, more interestingly, the manipulation of its FE polarization at room temperature is of great interest for next generation non-volatile memory devices with low power consumption. The control of the FE polarization  could be used as a knob for controlling the spin-to-charge current conversion sign \cite{varotto2021room}. Using fs light pulses to manipulate the FE polarization state in \hbox{$\alpha$-GeTe(111)} thus emerges as an exciting perspective for a drastic increase of the performance of the future generation of spintronics devices. Here, using fs extreme-ultraviolet pulses delivered by a table-top high-harmonic-generation (HHG) source \cite{puppin2019time}, we perform tr-ARPES as schematically depicted in Fig. \ref{fig1}a. It is a direct and comprehensive technique to track the momentum and energy resolved dynamical evolution of the electronic band structure which allows us to reveal a room temperature photoinduced enhancement of the Rashba coupling that directly signifies an increase of the ferroelectricity in \hbox{$\alpha$-GeTe(111)} on the subpicosecond timescale.

\bigskip

Figure \ref{fig1}b presents the band structure of thin films of \hbox{$\alpha$-GeTe(111)} along the $\bar{\Gamma}-\bar{K}$ high-symmetry direction, revealed by static ARPES measurements using only the probe pulses, in agreement with existing literature\cite{di2013electric,liebmann2016giant,krempasky2016disentangling,PhysResearchKrempasky2020,kremer2020,krempasky2021PRL}.  The observed band structure is in excellent agreement with Bloch spectral function (BSF) calculations (Fig. \ref{fig1}c) performed for a Te terminated surface with short bonds between the last two atomic lattice planes of the top Te surface and the next subsurface Ge plane (right part of Fig. \ref{fig1}a). In the vicinity of the Fermi level ($E_{F}$), the Rashba-split bulk states $B_{1}$ and $B_{2}$ are well reproduced. Similarly, two surface-derived states labelled $SS_{1}$ and $SS_{2}$ are well-resolved. Our surface is well ordered with Te termination and short surface Te--Ge bonds, as evidenced by the surface state dispersion\cite{rinaldi2018ferroelectric}, a crucial point for the rest of the discussion. This particular termination leads to a strong outward FE polarization which is at the origin of the large Rashba splitting of the bulk states. 

\bigskip

%---FIGURE 1 -------------Static---------------------------
%----------------------------------------------------------

\begin{figure}[H]
\begin{center}
\includegraphics[width=100mm]{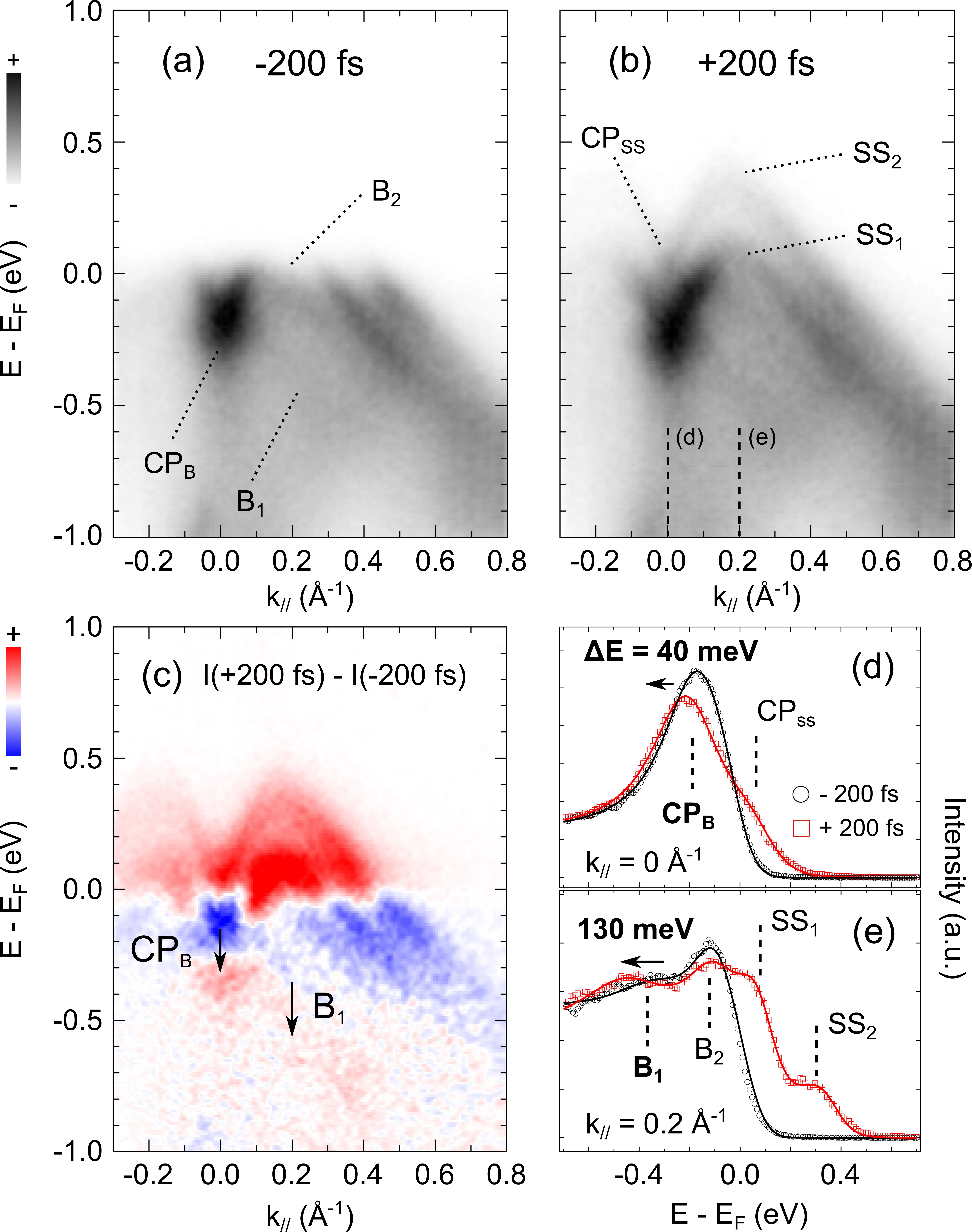}
\end{center}
\caption{\label{fig2} \textbf{Out-of-equilibrium tr-ARPES measurements.} (a) tr-ARPES spectra of $\alpha$-GeTe(111) recorded at pump-probe delays of (a) $-200$ fs and (b) $+200$ fs. The data have been acquired along the $\bar{K}-\bar{\Gamma}-\bar{K}$ direction with an absorbed pump fluence of 1 mJ/cm$^{2}$. (c) Difference plot between panels (b) and (a). Red and blue colors correspond to an increase and a depletion of photoemission intensity, respectively. The two arrows highlight the transient shift of the crossing point of the bulk states at normal emission (CP$_{B}$) and the B$_{1}$ contribution. This shift is visible as a red/blue contrast on the difference map. Energy distribution curves (EDCs) at $-200$ fs (black) and \hbox{$+200$ fs} (red) taken at (d) $k = \thinspace 0 \thinspace $\AA$^{-1}$ (normal emission) and (e) $k = \thinspace 0.2 \thinspace $\AA$^{-1}$ (off-normal emission) : see vertical dashed lines in panel (b). Solid lines correspond to a fit of the raw data.}
\end{figure}

We next investigate the out-of-equilibrium ultrafast dynamics of $\alpha$-GeTe(111) using a stroboscopic pump-probe approach. To do so, we excite the material with infrared (IR) pulses ($1.55$ eV pump pulses) and examine the response of the system by probing its electronic band structure with the $21.7$ eV probe pulses. Figure \ref{fig2}a,b shows the ARPES maps taken at time delays of $-200$ and $+200$ fs, respectively. The most striking observation is the transient population of the Rashba-split surface state $SS_{1}$ and $SS_{2}$ above $E_{F}$, which perfectly follows the transient elevation of the electronic temperature and the related Fermi-Dirac distribution broadening (see Fig. S2). This is confirmed by looking at the difference ARPES map in \hbox{Fig. \ref{fig2}c.}

\bigskip

%---FIGURE 2 -------------Dynamics---------------------------
%----------------------------------------------------------

The red color above $E_{F}$ corresponds to an increase of the photoemission intensity whereas the blue is associated to a decrease, confirming the transient Fermi-Dirac distribution broadening on this timescale and the thermal population of the surface state. Close inspection of the difference map in Fig. \ref{fig2}c reveals an additional transient feature. Indeed, a blue/red contrast is also visible in the bulk states region, approximatively \hbox{$400$ meV} below $E_{F}$. Here, the red shading (increase) is located at higher binding energy compared to blue ones (decrease), meaning that this intensity change is not related to the above mentioned Fermi edge broadening. Instead, it rather corresponds to a transient shift of the bulk states to high binding energy. We evidence this in \hbox{Fig. \ref{fig2}d,e} with energy distribution curves (EDCs) taken at \hbox{$k = \thinspace 0 \thinspace $\AA$^{-1}$} (normal emission) and \hbox{$k = \thinspace 0.2 \thinspace $\AA$^{-1}$} (off-normal emission). They show a transient shift of the crossing point of the bulk states (CP$_{B}$) and $B_{1}$ contribution after $200$ fs, respectively, equal to $40$ and $130$ meV (the $B_{2}$ contribution does not shift). Since the Rashba splitting is proportional to the energy difference between the $B_{1}$ and $B_{2}$ branches, this is a clear evidence for a transient light-induced increase of the Rashba coupling in $\alpha$-GeTe(111) on the sub-picosecond (ps) timescale. 

\bigskip

\begin{figure}[H]
\begin{center}
\includegraphics[width=100mm]{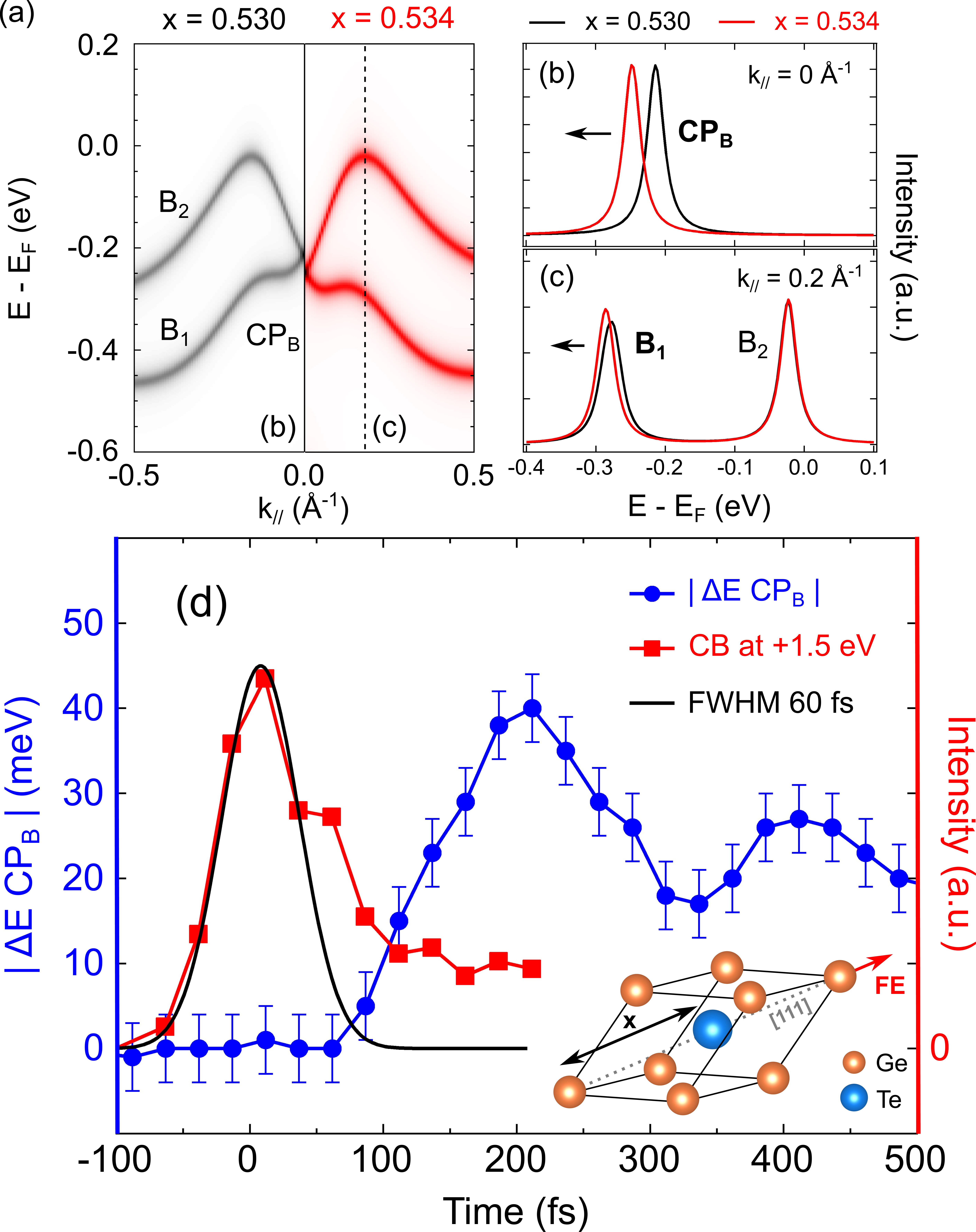}
\end{center}
\caption{\label{fig3} \textbf{Band structure calculation and temporal evolution of CP$_{B}$ position.} (a) Bulk band structure of $\alpha$-GeTe(111) along the $\bar{\Gamma}-\bar{K}$ high-symmetry direction for two different x positions of the Te atom, as defined in the inset of panel (d). Corresponding EDCs for $x \thinspace = \thinspace 0.530$ (black) and $x \thinspace = \thinspace 0.534$ (red) taken at (b) $k = \thinspace 0.0 \thinspace $\AA$^{-1}$ and (c) $k = \thinspace 0.2 \thinspace $\AA$^{-1}$. (d) Temporal evolution of CP$_{B}$ extracted from tr-ARPES measurements taken at an absorbed pump fluence of $1.5$ mJ/cm$^{2}$ (blue curve). The red curve corresponds to the transient population of the conduction band and the black curve to the corresponding Gaussian fit.}
\end{figure}

To explain the structural origin of our observation in the electronic structure, Fig. \ref{fig3}a shows band structure calculations around the $\bar{\Gamma}$ point of the BZ for two distinct FE distortions, where we only show the bulk contributions for the sake of clarity. The left panel corresponds to the ground state of the system where the x position  of the Te atom is 0.530, expressed as a fraction of the distance between two Ge atoms along the [111] direction (see Fig. \ref{fig3}d inset). Alternatively, the right panel in Fig. \ref{fig3}a shows the case where the FE distorsion has increased  to x $\thinspace = \thinspace 0.534$ which corresponds to a $13  \thinspace \%$ increase in the FE polarization. As evidenced by Fig. \ref{fig3}b,c, such an increase of ferroelectricity leads to a shift of CP$_{B}$ at the $\bar{\Gamma}$ point and, at the same time, for higher momenta to bulk band shift B$_{1}$ to higher binding energy. This is in good qualitative agreement with our experimental observations shown in Fig. \ref{fig2} and consequently confirms our interpretation since it reproduces well the 40 meV shift of the CP$_{B}$ we experimentally observe. As x approaches $0.5$, the system tends towards a more centrosymmetric structure  (rock salt structure, space group $Fm\bar{3}m$) and consequently to the more paralectric phase with a less important Rashba coupling (see Fig. S3). Consequently, the ferroelectricity transiently increases, which is highly unusual, as photoexcitation typically leads to a more symmetric phase\cite{huber2014coherent,nicholson2018beyond}. 

\bigskip 

%---FIGURE 3 -------------Calculations---------------------------
%----------------------------------------------------------

As a matter of fact, the energy position of CP$_{B}$ is proportional to the magnitude of the FE distortion in \hbox{$\alpha$-GeTe(111).} To obtain the temporal dynamics of the ferroelectricity induced by photoexcitation, we consequently plot the time evolution of the CP$_{B}$ shift, \hbox{$\Delta E$ CP$_{B}$}, as extracted from our tr-ARPES measurements. This evolution displays two remarkable dynamical features. Firstly, it takes about $80$ fs before a CP$_{B}$ shift occurs. The position of time zero has been obtained from the transient electronic population $1.5$ eV above $E_{F}$ in the conduction band (red curve). Secondly,  after $200$ fs the CP$_{B}$ shift reaches its maximum of $40$ meV, followed by a coherent oscillation with a period of approximatively $200$ fs, corresponding to a frequency of $5$ THz (see also Fig. S4). Thus, the effect of the IR pump pulse is to photoinduce on the sub-ps timescale a coherent modulation of the ferroelectricity in $\alpha$-GeTe(111). Due to the coupling of Bloch states to this FE $A_{1g}$ mode along the [111] direction of the crystal \cite{dangic2021origin}, we thereby observe a resultly modulation of the Rashba coupling as probed by our tr-ARPES measurements. 

\bigskip

The observations presented above are in notable disagreement with previous literature. Indeed, experimental works using ultrafast electron diffraction\cite{hu2015transient} and time-resolved X-ray diffraction\cite{matsubara2016initial} observed instead that an IR pump pulse produces a successive FE to paraelectric, and a paraelectric to amorphous phase transition in $\alpha$-GeTe. These results were supported by time-dependent DFT calculations suggesting that even in the lowest fluence regime, the effect of an IR pulse on $\alpha$-GeTe is a diminuation of the ferroelectricity caused by a reduction of the Ge--Te bonds length\cite{Chen2018,Chen2020}. Last but not least, we observe an unexpected delay of $80$ fs before the coherent oscillation sets in. All these observations are unusual in a standard displasive excitation of a coherent phonon (DECP) picture.

\bigskip

\begin{figure}[H]
\begin{center}
\includegraphics[width=160mm]{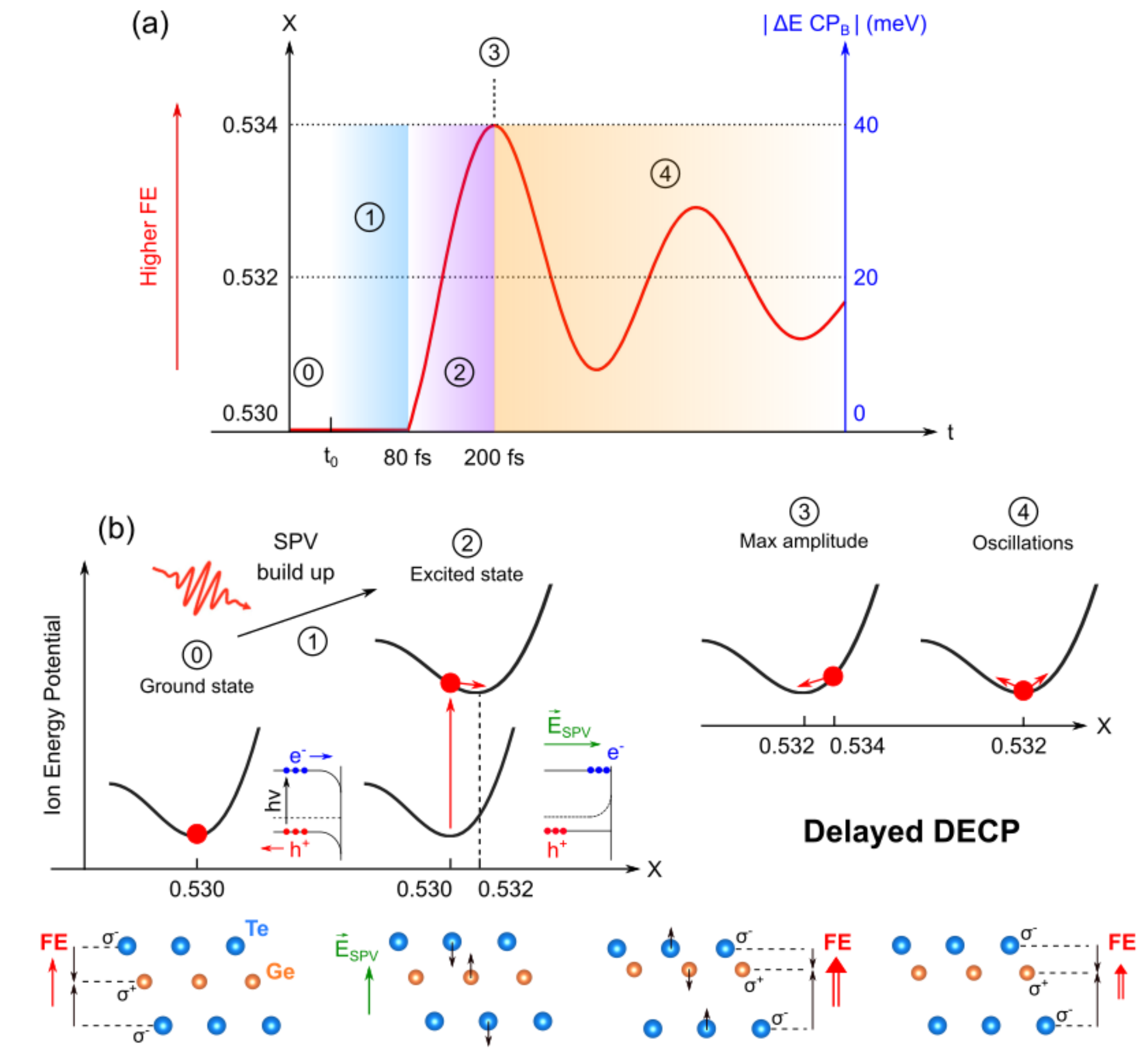}
\end{center}
\caption{\label{fig4} \textbf{Illustration of the delayed DECP mechanism.} (a) Schematic evolution of the temporal evolution of the x position of the Te atom in the primitive unit cell of $\alpha$-GeTe(111) and the corresponding variation of the position of CP$_{B}$ as measured with ARPES. The five steps of the mechanism are numbered from \textcircled{0} to \textcircled{4}. (b) Evolution of the subsurface structure and potential of $\alpha$-GeTe(111) during these five steps. The IR pulse produces a SPV leading to a transient increase of the lattice distorsion and consequently of the ferroelectricity.}
\end{figure}

 To understand this discrepency, we have to consider that ARPES probes mainly the surface of $\alpha$-GeTe(111), unlike the bulk sensitive previous experimental studies cited above. Figure \ref{fig4}a summarizes the most important steps of the temporal evolution of the ferroelectricity in $\alpha$-GeTe(111) after photoexcitation, which scales with the x position of the Te atom along the $\alpha$-GeTe  [111] crystallographic direction (vertical black axis). At the same time the transient shift of the CP$_{B}$ energy position is displayed on a blue vertical axis, which relates to the experimental blue lineout in Fig. \ref{fig3}d. The five steps of the microscopic mechanism behind the dynamics evolution is detailed in Fig. \ref{fig4}.  \textcircled{0} At negative delays, we probe the ground state of the system where the minimum of the ion potential corresponds to x $= 0.530$. \textcircled{1} Within the first $80$ fs after photoexcitation, a surface photovoltage (SPV) builds up and modifies the electronic energy level diagram of the system. SPV is a well known phenomenon in semiconductors physics\cite{zhang2012band,BPkremer}: it corresponds to a photoinduced spatial redistribution of charge carriers and to a compensation of the pre-existing band bending (BB) in the surface charge region (SCR). Evidence of a positive SPV in our case is given in Fig. S5, where we show the pump induced shift to high kinetic energy of the whole ARPES spectrum. As explained in \hbox{Fig. S6,} a positive SPV corresponds to a migration of electrons to the surface and holes to the bulk, creating a compensative electric field in the direction out of the sample surface. The SPV build up typically takes a few tens to a few hundreds of fs\cite{zhang2012band,BPkremer}. We performed drift-diffusion calculations on a GeTe surface (see Fig. S7), demonstrating that the pump-pulses completely suppress the initial BB after a few tens of fs, in excellent agreement with our experimental findings.   \textcircled{2} The compensative electric field plays an additionnal and crucial role since it influences the subsurface ferroelectricity. As depicted in this second step of Fig. \ref{fig4}b, the Te atoms are pushed into the bulk and Ge atoms towards the surface, leading to an increase of the FE distortion. Thus, the system is driven into a transient out-of-equilibrium state in which the new ion energy minimum is shifted to higher x with respect to the ground state. \textcircled{3} After $200$ fs, the ion displacement reaches the maximum amplitude corresponding to x $= 0.534$, associated to the $40$ meV shift of CP$_{B}$, as evidenced by our tr-ARPES measurement at $+200$ fs. \textcircled{4} Finally, the system oscillates around its new potential minimum localized at an intermediate position of x $= 0.532$ with concomitant periodic oscillation of the ferroelectricity that correlates with the CP$_{B}$ position and Rashba splitting. As a whole, these five steps explain the experimental behaviours and appear as a delayed DECP mechanism.

\bigskip

%---FIGURE 4 -------------MEchanism---------------------------
%----------------------------------------------------------

In this picture, we understand the light-induced modulation of ferroelectricity in $\alpha$-GeTe(111) as a surface phenomenon. It is compatible with the rapid fluence saturation of the CP$_{B}$ shift that we observe (see Fig. S8), since it is known that the SPV typically saturates in a very low fluence regime \cite{chen2020spectroscopy}. The non-negligible SPV build up time is also consistent with the delay we observe in the CP$_{B}$ shift. 

\bigskip

To conclude, we have demonstrated ultrafast light induced modulation of ferroelectricity in the Rashba semiconducting material $\alpha$-GeTe(111) with tr-ARPES measurements. We observed a remarkable delayed enhancement of the Rashba coupling that can be understood by invoking a delayed DECP mechanism originating from the creation of an SPV on the fs timescale inside the sub-surface region. Beyond its implications for the fundamental research adressing the out-of-equilibrium dynamics of semiconducting ferroelectric materials, our study validates the use of fs light pulses for the manipulation of the polarization in $\alpha$-GeTe(111) at room temperature and opens promising new routes for technological applications in spintronics, especially for memory devices. Systematic experiments using variable photoexcitation wavelength, for example in the THz regime, are desired to optimize the amplitude of the photoinduced polarization enhancement beyond our proof-of-principle experiment. Finally, given that this effect is confined to the surface region, it should also be effective in the ultrathin limit of a few nanometers, an appealing perspective given the current interest in heterostructure materials.

\section{Methods}

\textbf{Time resolved photoemission spectroscopy.} The samples were transferred to the tr-ARPES setup using an ultra high vacuum (UHV) suitcase  with a base pressure $<  1  \thinspace \times \thinspace  10^{-10}$  mbar. All tr-ARPES measurements were performed in UHV  at P $< 1  \thinspace \times \thinspace  10^{-10}$  mbar and at room temperature, using a laser-based high-harmonic-generation trARPES setup \cite{puppin2019time} (\hbox{h$\nu_{probe}$ = 21.7  eV}, h$\nu_{pump}$ = 1.55  eV, 500  kHz repetition rate, $\Delta E \thinspace \approx \thinspace $ 175  meV, $\Delta t \approx \thinspace $ 35 fs) with a SPECS Phoibos 150 hemispherical analyzer and a 6-axis manipulator (SPECS GmbH). The pump and probe spot sizes (FWHM) are $\approx \thinspace 150 \thinspace \times \thinspace 150  \thinspace \mu m^{2}$ and $\approx \thinspace 70 \thinspace \times \thinspace 60  \thinspace \mu m^{2}$, respectively. All discussed fluences refer to the absorbed fluence, determined using the complex refractive index \hbox{$ n = \sqrt{\epsilon} \approx 5.5 + 4.5 \thinspace i$} at $\lambda  = 800 $ nm. 

\newpage

\noindent \textbf{Computational details.} 

\noindent \textit{Electronic structure calculations -} The presented ab-initio calculations are based on fully relativistic density functional theory as implemented within the multiple scattering Korringa-Kohn-Rostoker Green function based package (SPRKKR)\cite{ebert2011calculating}. Relativistic effects such as spin-orbit coupling are treated by Dirac equation. The LDA was chosen to approximate the exchange-correlation part of the potential along with the atomic sphere approximation. The electronic structure is represented using the Bloch Spectral Function (BSF) which consists of the imaginary part of the Green function. In the case of Fig. \ref{fig1}c, a semi-infinite crystal of $\alpha$-GeTe(111) with Te surface termination was considered, as in our previous work\cite{kremer2020}, representing electronic states of both bulk and surface natures. For Fig. \ref{fig3}a-c, an infinite crystal is here treated, accounting therefore only for electronic states of bulk character.

\noindent \textit{Drift-diffusion calculations -} The drift-diffusion equations have been solved with a custom code as described in Ref.\cite{BPkremer} using the parameters listed in the Extended data. 

\bigskip

\noindent \textbf{Samples growth.} Ferroelectric $\alpha$-GeTe films were grown by molecular beam epitaxy on (111) oriented BaF$_2$ substrates using Ge and Te effusion cells and a growth temperature of 280$^{\circ}$C. Perfect 2D growth was observed by in situ reflection high-energy electron diffraction. After growth, the samples were transferred into a Ferrovac UHV suitcase in which they were transported for the tr-APRES measurements without breaking UHV conditions \hbox{($<  1  \thinspace \times \thinspace  10^{-10}$  mbar).} 

\section{Acknowledgements}

G.K, C.W.N. and C.M. acknowledge support from the Swiss National Science Foundation  Grant \hbox{No. P00P2$\_$170597}. Ja.M. and L.N. would like to thank the CEDAMNF (Grant No. CZ.02.1.01/0.0/0.0/15$\_$003/0000358) co--funded by the Ministry of Education, Youth and Sports of Czech Republic and the GACR Project No. 20-18725S for funding. This work was supported by the Max Planck Society, the European Research Council (ERC) under the European Union’s Horizon 2020 research and innovation program (Grant No. ERC-2015-CoG-682843), and the German Research Foundation (DFG) within the Emmy Noether program (Grant No. RE3977/1). C. Y. and P. W. acknowledge support by SNSF Grant No. 200021-196966. The drift-diffusion calculations have been performed on the Beo05 cluster at the University of Fribourg. G.S. acknowledges support by the Austrian Science Funds, Projects P30960-N27 and I4493-N. 

\section{Contributions}

J.K., G.K. and C.M. conceived the project. G.S. prepared the samples. Tr-ARPES measurements were performed by G.K., Ju.M., C.W.N., L.R. and C.M. with the help of L.N. and C.S., and were analysed by G.K.. Electronic structure calculations were carried out by L.N. and Ja.M.. The drift-diffusion calculations have been performed by C.Y. and P.W..  The project was coordinated by G.K. together with C.M.. All the authors participated to the interpretation of the data. G.K. wrote the manuscript with input from all authors. L.N. and C.W.N. equally contributed to this work.

\section{Corresponding authors}
*E-mail: geoffroy.kremer@universite-paris-saclay.fr

\noindent **E-mail:  jminar@ntc.zcu.cz

\section{Competing interests}
The authors declare no competing interests.

\newpage

\bibliography{biblio}



\section*{Supplementary material}

# Extended data for : Field-induced ultrafast modulation of Rashba coupling at room temperature in ferroelectric $\alpha$-GeTe(111)

G. Kremer,[†,‡] J. Maklar,[¶] L. Nicolaï,[§] C. W. Nicholson,[†,¶] C. Yue,[†] C. Silva,[¶] P. Werner,[†] J. H. Dil,[∥,⊥] J. Krempaský,[∥] G. Springholz,[#] R. Ernstorfer,[¶,@] J. Minár,[§] L. Rettig,[¶] and C. Monney[†]

[†]Département de Physique and Fribourg Center for Nanomaterials, Université de Fribourg, CH-1700 Fribourg, Switzerland

[‡]Université Paris-Saclay, CNRS, Centre de Nanosciences et de Nanotechnologies, 91120, Palaiseau, France

[¶]Fritz Haber Institute of the Max Planck Society, Faradayweg 4-6, 14195 Berlin, Germany

[§]New Technologies-Research Center University of West Bohemia, Plzen, Czech Republic

[∥]Photon Science Division, Paul Scherrer Institut, CH-5232 Villigen, Switzerland

[⊥]Institute of physics, Ecole Polytechnique Fédérale de Lausanne, CH-1015 Lausanne, Switzerland

[#]Institut für Halbleiter-und Festkörperphysik, Johannes Kepler Universität, A-4040 Linz, Austria

[@]Institut für Optik und Atomare Physik, Technische Universität Berlin, Straße des 17, Juni 135, 10632 Berlin, Germany

E-mail:

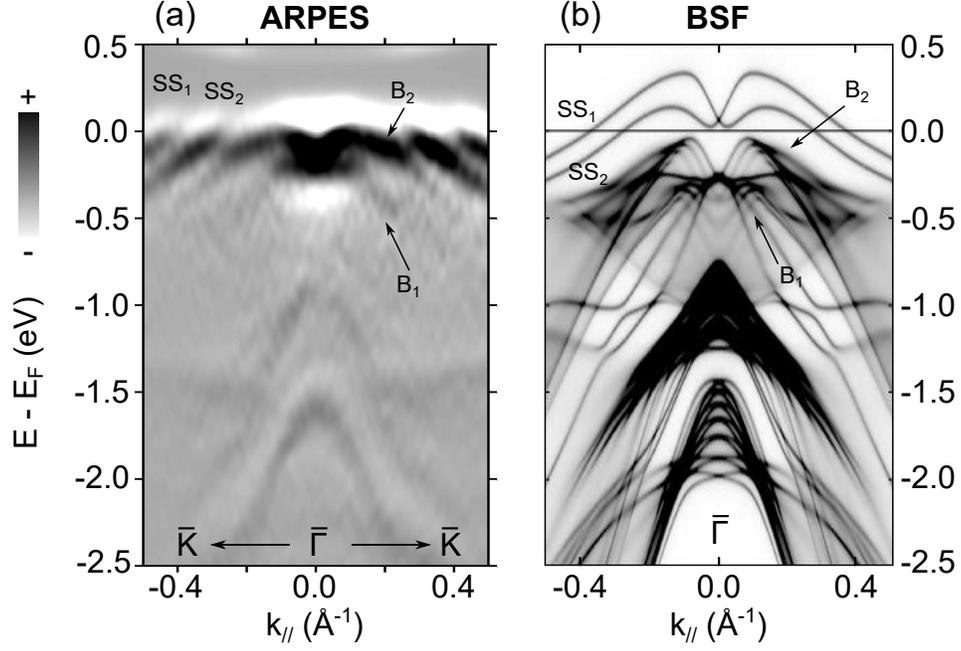

Figure S1: (a) Second derivative of static ARPES spectrum along the $\bar{K} - \bar{\Gamma} - \bar{K}$ high-symmetry direction and (b) corresponding BSF calculations for a Te-terminated surface of $\alpha$-GeTe(111) with short surface bonds.

## Second derivative of static $\alpha$-GeTe(111) ARPES spectrum

Figure S1(a) shows the second derivative of ARPES spectrum in Figure 1(b) in the main text. It better highlights the dispersion of Rashba-split bulk branches $B_1$ and $B_2$.

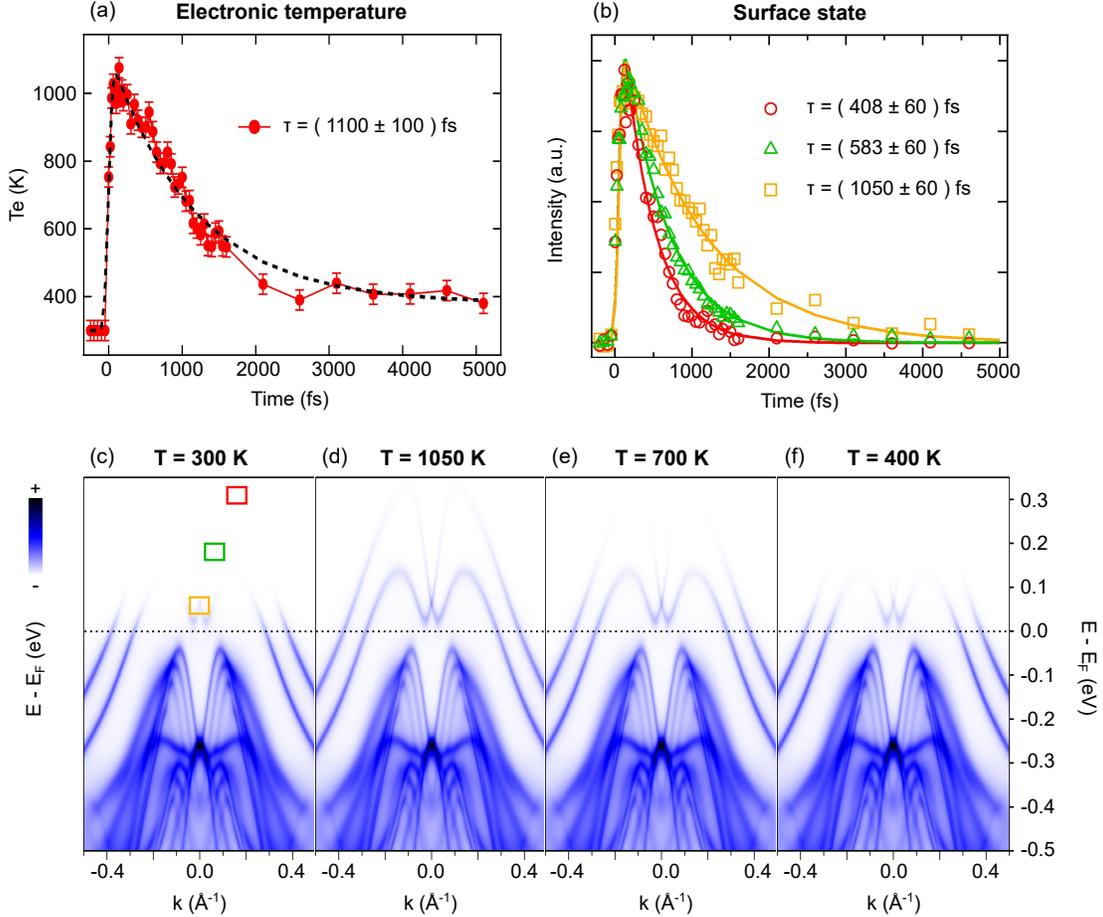

Figure S2: (a) Transient electronic temperature extracted by fitting the Fermi-Dirac distribution in the experimental data at different time delays for an absorbed fluence of 0.5 mJ/cm$^2$. (b) Transient photoemission intensity from experimental data integrated within the colored integration squares indicated in panel (c).(c) BSF band structure calculations of $\alpha$-GeTe(111) multiplied by Fermi-Dirac distribution at different temperatures obtained from panel (a).

## Thermal surface state population

Figure S2(a) shows the temporal evolution of the transient electronic temperature as extracted by fitting the Fermi-Dirac distribution in the experimental data. It shows a maximum at 180 fs corresponding to $T_e = 1050$ K before slowly recovering with a time constant of 1.1 ps, as fitted with an exponential decay. This is a standard behaviour for a semiconducting material.[1] In Figure S2(b) we plot the transient evolution of the photoemission intensity at different positions along the surface state dispersion. The higher we are above $E_F$, the faster is the recovery time constant extracted from an exponential decay. Nevertheless, the rise

3

time is the same for each curve and their maximum correspond to 180 fs. This can be understood by plotting the calculated surface state dispersion convoluated by the Fermi-Dirac distribution at different temperatures, as shown in Figure S2(c). The higher the temperature is, the more the surface state is populated above $E_F$. So we can conclude that the surface state is populated above $E_F$ due to the transient increase of the electronic temperature. This is confirmed by comparing the orange curve in panel (b) and the electronic temperature from panel(a). The two curves perfectly fit.

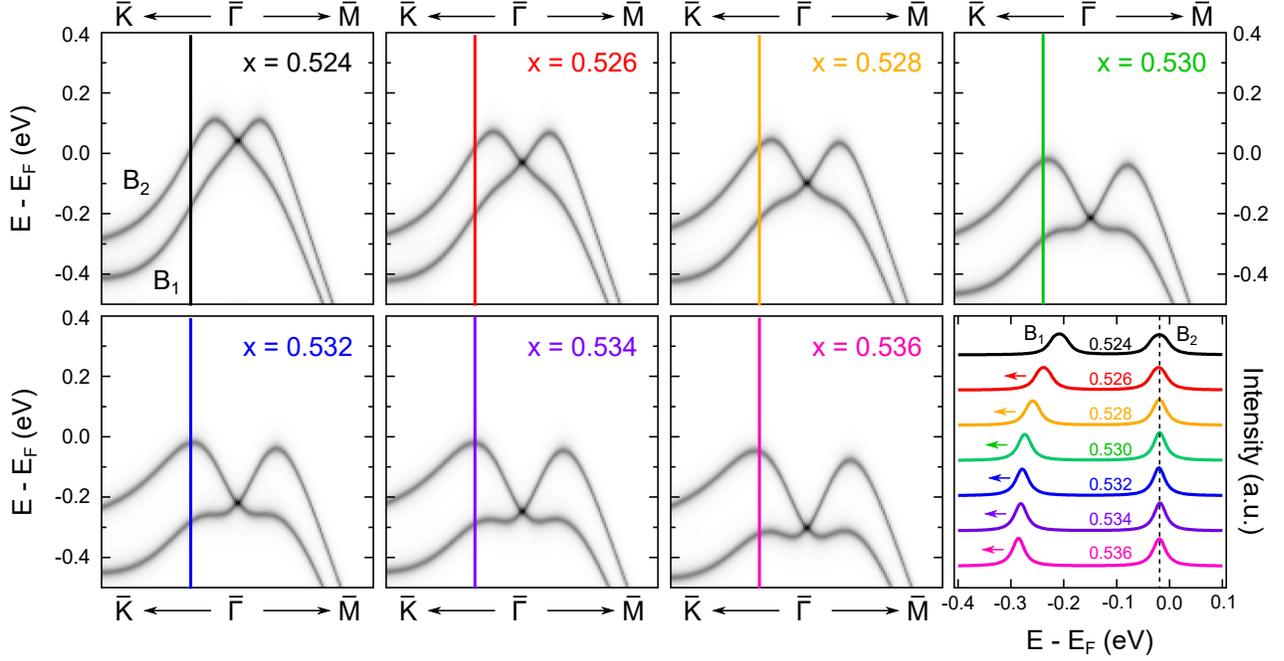

Figure S3: Band structure calculation of $\alpha$-GeTe(111) for a series of x positions of the Te atom in the primitive unit cell, as defined in the inset of Figure 3d. The ground state corresponds to x = 0.530. Bottom right panel shows EDCs aligned relatively to the $B_2$ contribution away from the $\bar{\Gamma}$ point (see vertical colored lines in other panels).

## Band structure calculations as a function of the ferroelectric distorsion

Figure S3 shows band structure calculations of $\alpha$-GeTe(111) for a series of x positions of the Te atom in the primitive unit cell. The more the value of x deviates from 0.5, corresponding to the centrosymmetric paraelectric phase, the more the Rashba splitting is important. This is illustrated in the bottom right panel. The $B_2$ contribution has been fixed to the position for x = 0.524 and the $B_1$ contribution shifts to higher BE when x is increasing.

5

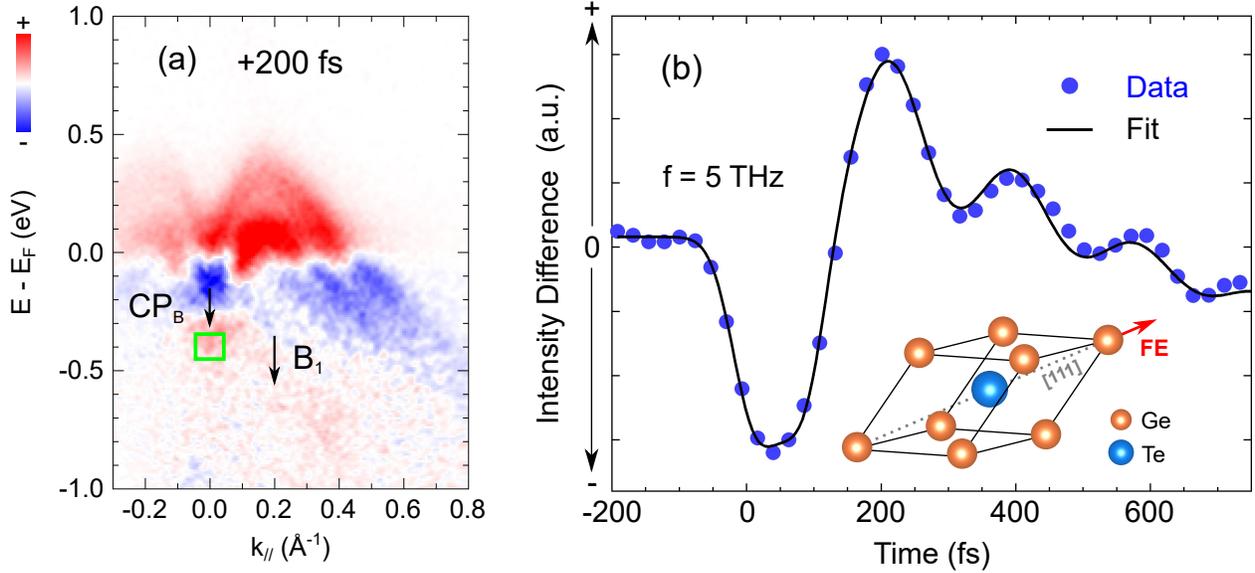

Figure S4: (a) +200 fs difference tr-ARPES intensity map. (b) Transient photoemission intensity within the green integration square indicated in panel (a) showing a coherent oscillation at a frequency of 5 THz.

## Coherent $A_{1g}$ phonons mode

Figure S4(a) shows the difference tr-ARPES map at +200 fs, corresponding to the maxium amplitude of delayed DECP mechanism as explained in the main text. In Figure S4(b) we plot the transient photoemission intensity within the green box in panel (a). It shows a coherent modulation with a 5 THz frequency as extracted from fitting procedure, in good agreement with our extraction of the transient evolution of the $CP_B$ shift discussed in Figure 3 of the main text. This coherent modulation is associated to the $A_{1g}$ phonon mode, as evidenced by bulk phonon dispersions calculations at $\bar{\Gamma}$ in the FE phase.[2] At this high symmetry point a 4.4 THz frequency value is reported. The deviation with our measurements can be explained both by the difference of the subsurface structure with the bulk one and by the transient modification of the shape of the ion energy potential landscape.

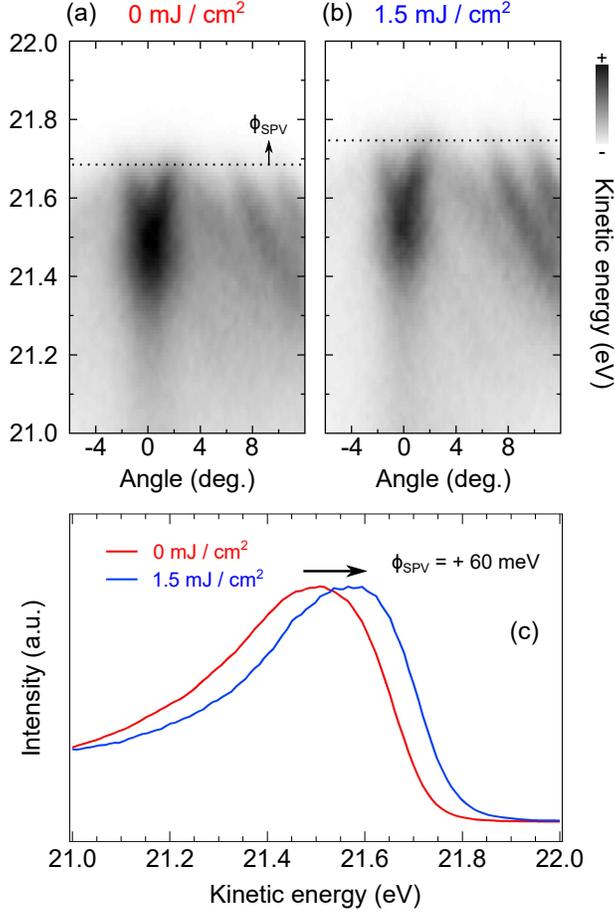

Figure S5: tr-ARPES measurements of $\alpha$-GeTe(111) at negative pump-probe delay for an absorbed fluence of (a) 0 mJ/cm$^2$ and (b) of 1.5 mJ/cm$^2$. (c) Corresponding angle integrated EDCs. A positive SPV of +60 meV is observed due to the compensation of the downward BB as depicted in Figure S6. The kinetic energy has been corrected by the work function of the photoemission analyser.

## Evidence of a positive surface photovoltage (SPV)

Figure S5 displays the tr-ARPES measurements of $\alpha$-GeTe(111) at $-1$ ps pump-probe delay without (red) and with (blue) the pump. From this, it is possible to evaluate the magnitude and the sign of the SPV.[3,4] We observe that the blue spectrum is shifted by +60 meV to high kinetic energy with respect to the red one. A downward band bending (BB) is expected for p doped $\alpha$-GeTe(111) with surface donor states, as illustrated in the top panel of Figure S6(b). This is confirmed by the positive value of the SPV which is compensating the downward BB.

7

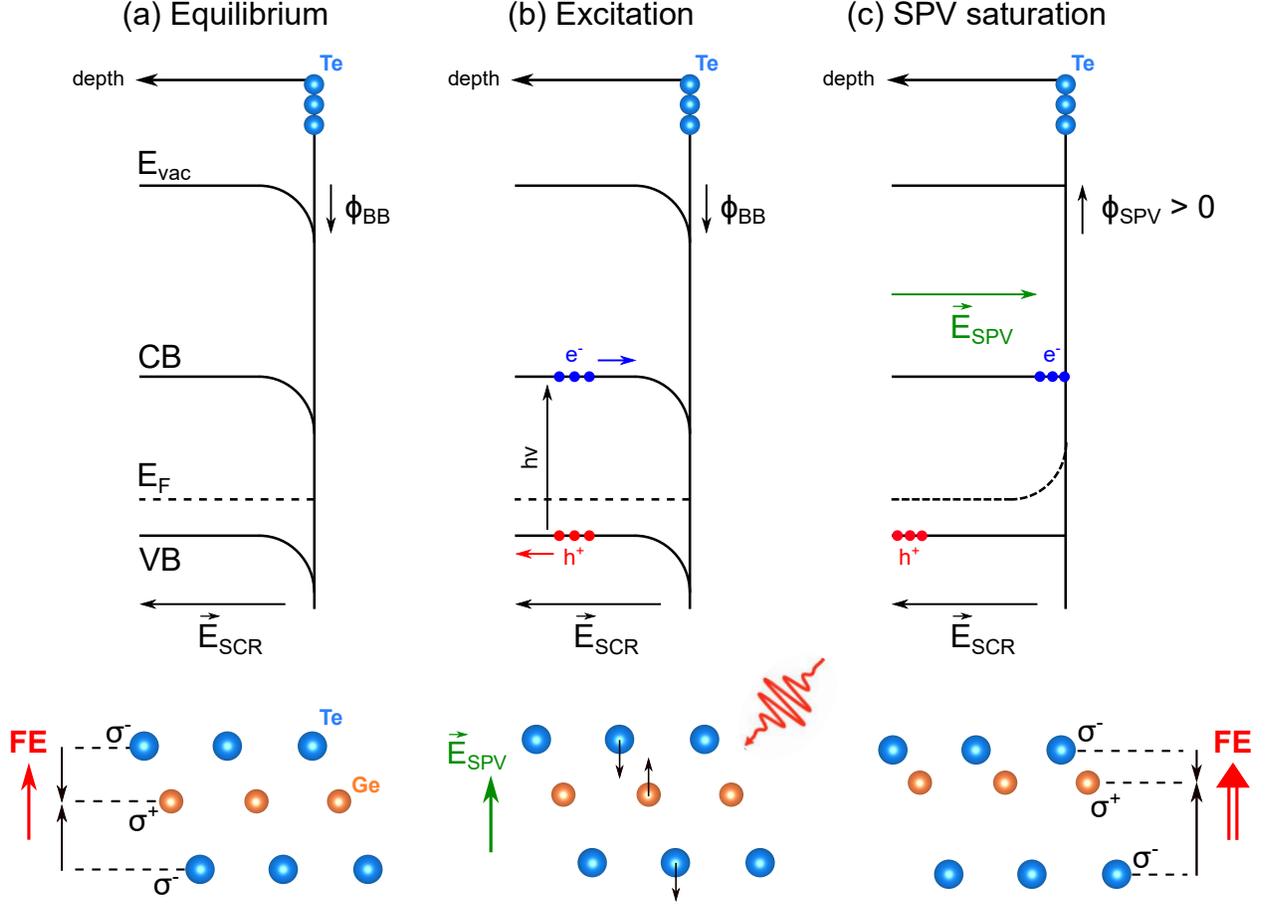

Figure S6: Schematic diagram of electron energy levels near the surface (top) and corresponding subsurface structure (bottom) of $\alpha$-GeTe(111): (a) At the equilibrium, (b) under the illumination and (c) after the charge carriers redistribution and SPV saturation.

## Downward BB and positive SPV in $\alpha$-GeTe(111)

Figure S6 displays the energy band diagrams of a p-doped $\alpha$-GeTe(111) surface. In each case, we consider the (a) nonequilibrium, (b) the equilibrium and the (c) photoexcited configurations. For a p-doped surface with surface donors states, a downward BB and consequently a positive SPV are expected. These expectations are confirmed by our measurements in Figure S5. In the particular case of a FE material, the generated compensative electric field ($E_{SPV}$) affects the initial surface dipole. In the present case, we expect that photoinduced $E_{SPV}$ push Te atoms into the bulk and Ge atoms into the surface, leading to a photoinduced increase of the FE distorsion: see bottom panels. In other words, $E_{SPV}$ reinforces the ini-

8

tial net dipole at the surface originating from the Te surface termination and short bonds between the first Te and the subsurface Ge planes.

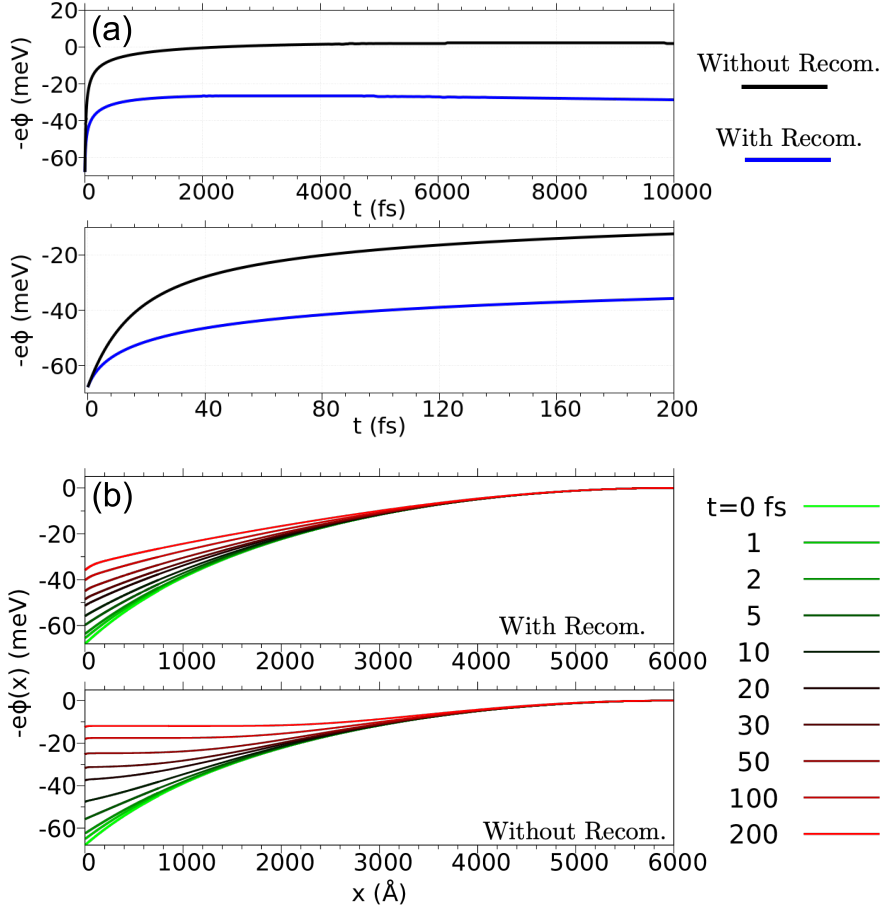

Figure S7: (color online). Non-equilibrium dynamics obtained by solving the drift-diffusion equations. (a) The SPV as a function of time after the pump for systems with recombination (blue) and without recombination (black). (b) The band energy as a function of $x$ at different times. The upper (lower) panel is for system with (without) recombination. The parameters used in the simulation are shown in Tab. 1.

## Simulation of the carrier dynamics in photoexcited GeTe

We apply the numerical procedure presented in Ref[4] to solve the drift-diffusion equations (DDEs) in combination with the Poisson equation, using the realistic parameters listed in Tab. 1.

Let us first clarify the static situation before photo-doping. Since the surface states have a higher energy than the bulk states, they can dope electrons into bulk, which leaves holes confined in the surface states. We assume that the holes remain trapped near the surface with an exponentially decaying density profile $N_d(x) = n_u \lambda_d e^{-\lambda_d \frac{x}{l_0}}$, where $l_0 = 1$ is the unit

10

Table 1: Parameters used in the simulation. The diffusion coefficients are obtained by the Einstein relation $D_{n,p} = \frac{\mu_{n,p} k_B T}{e}$. The values of $\mu_n$, $\mu_p$, $\varepsilon_r$ are from Ref.[5,6] The coefficient $\alpha$ in $B_r$ is a unit-less number. We choose $\alpha = 0$ and $\alpha = 0.3$ in the simulation.

| Name | Symbol | Value |
|---|---|---|
| Temperature | $T$ | 300K |
| Relative Permittivity | $\varepsilon_r$ | 36.0 |
| Electron Mobility | $\mu_n$ | $100.0 \times 10^{-4} \text{m}^2/(\text{Vs})$ |
| Electron Diffusion Coefficient | $D_n$ | $2.585 \times 10^{-4} \text{m}^2/\text{s}$ |
| Hole Mobility | $\mu_p$ | $100 \times 10^{-4} \text{m}^2/(\text{Vs})$ |
| Hole Diffusion Coefficient | $D_p$ | $2.585 \times 10^{-4} \text{m}^2/\text{s}$ |
| Total surface density of holes | $\sigma_0$ | $8.0 \times 10^{14} \text{m}^{-2}$ |
| Photo-excited Carrier Volume | $\sigma_p$ | $\sim 1399\sigma_0$ |
| Length Unit | $l_0$ | $10^{-10}$m (1Å) |
| Sample Length | L | $6000 \cdot l_0$ |
| Number of $x$-grid points | N | 30001 |
| Electric Field Unit | $E_u = \frac{\sigma_0 e}{2\varepsilon_r \varepsilon_0}$ | $2.011 \times 10^5$V/m |
| Density Unit | $n_u = \frac{\sigma_0}{l_0}$ | $8.0 \times 10^{24} \text{m}^{-3}$ |
| Radiative Recombination Rate | $B_r = \frac{\alpha}{10\text{fs} \cdot n_u}$ | $1.25\alpha \times 10^{-11} \text{m}^3/\text{s}$ |

of length, $\sigma_0$ the total surface density of holes, $n_u = \sigma_0/l_0$ the unit of density. We choose $\lambda_d = 1/5$, which means the density of holes decays within 5Å to the bulk. We determine $\sigma_0 = 8.0 \times 10^{14} m^{-2}$ using the information on the band bending in equilibrium. The positive charge from this hole distribution will lead to an accumulation of electrons near the surface. However, because of the large relative permittivity $\varepsilon_r$ and large diffusion coefficients, the free electrons distribution $n(x)$ decays much more slowly into the bulk. Charge neutrality requires $\int N_d(x) = \int n(x) \equiv \sigma_0$. In static state, there are more holes near surface and more free electrons away from the surface. This electron-hole separation creates a space charge region, with a net electric field $E$ pointing to the bulk. We find the maximum value of $E \approx 2E_u$, with $E_u = 2.011 \times 10^5$V/m.

We assume that the photo-excited electron-hole pairs are generated instantaneously at $t = 0_+$ with the density profile $\delta n(x, t = 0_+) = \delta p(x, t = 0_+) = \gamma_0 e^{-x/\lambda_{ph}}$, where $\lambda_{ph} = 80$ nm is penetration depth and $\gamma_0$ the photo-excited carrier (surface) density with $\gamma_0 \approx 1399\sigma_0 \gg \sigma_0$. To take inter-band recombination into account, we introduce an ad-hoc term $B_r \delta_n \delta_p$. The

DDEs are solved for systems at $T = 300$K and the results are shown in Figure S7. The SPV $\phi_{\text{SPV}}$ as a function of $t$ is plotted in Figure S7(a), with the upper panel showing the time range $0 < t < 10000$ fs and the lower panel a zoom of the time range $0 < t < 200$ fs. Upon the photo-doping, we find a fast change in $\phi_{\text{SPV}}$ in first 10 fs, but it takes much longer time $\sim 100$ fs to reach half the maximum change in $\phi_{\text{SPV}}$. This is much slower than the time scale $\sim 1$ fs obtained for black phosphorus (BP),[4] since the drift mobility in GeTe is around $v = \mu E \sim 0.04$ Å/fs, which is nearly 250 times smaller than that ($\sim 10$ Å/fs) in BP. The corresponding energy band -e$\phi(x)$ is shown in Figure S7(b). These data show that the internal electric field is weakened in the photo-doped systems (flattening of the energy bands), which is much more evident in system without recombination.

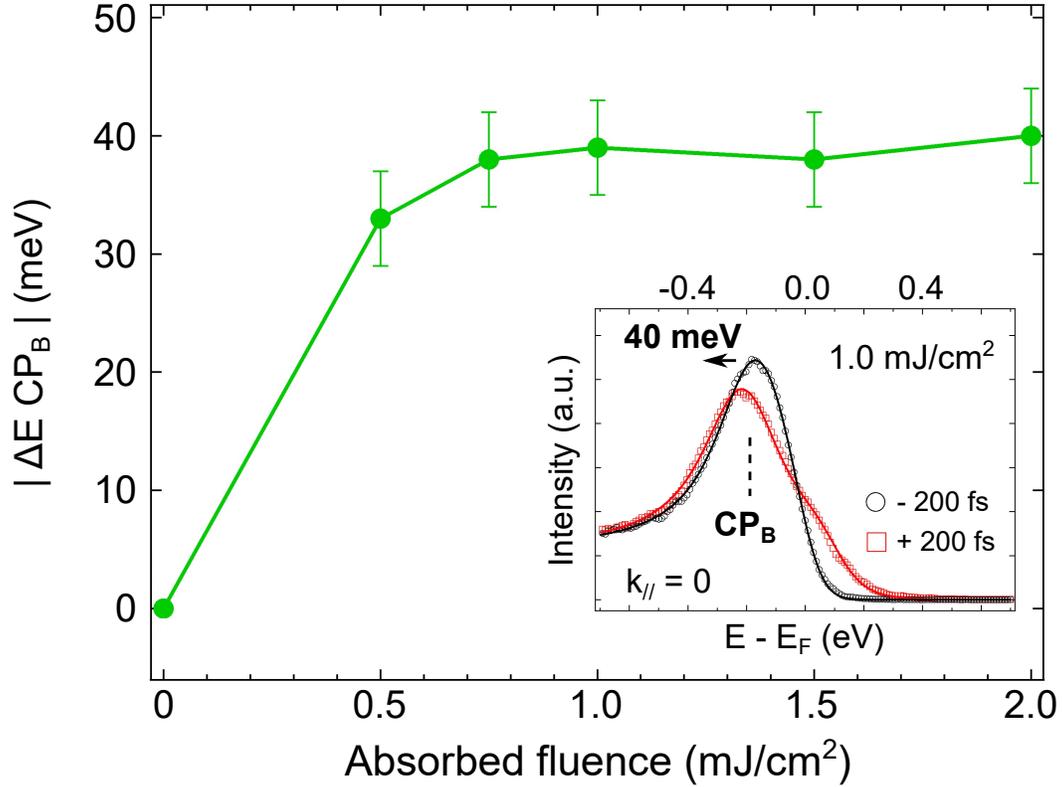

Figure S8: Absorbed fluence dependence of the photoinduced shift of the $CP_B$ at +200 fs as extracted from EDCs at the $\bar{\Gamma}$ point : see inset.

## Fluence dependence

Figure S8 displays the fluence dependence of the $CP_B$ at +200 fs. It shows a rapid saturation at a 0.5 mJ/cm$^2$ treshold. This is unexpected in a standard DECP picture but can be understood in our delayed DECP mechanism invoking SPV. Indeed, it is well documented that SPV quickly saturates as a function of the fluence.[7]



\end{document}